\documentclass[conference]{IEEEtran}
\IEEEoverridecommandlockouts
\usepackage{cite}
\usepackage{times} 
\usepackage{amsmath,amssymb,amsfonts}
\usepackage{algorithm}
\usepackage{graphicx}
\usepackage{textcomp}
\usepackage{amsmath, amssymb, amsthm}
\usepackage{bm}
\usepackage{enumitem}
\usepackage{algpseudocode}
\usepackage{hyperref}
\usepackage{tikz}          
\usepackage{xcolor}
\usepackage{setspace}      

\newcommand{\titlefont}{\fontsize{24pt}{30pt}\selectfont} 
\newcommand{\authorfont}{\fontsize{10pt}{12pt}\selectfont} 
\newcommand{\affilfont}{\fontsize{10pt}{12pt}\selectfont}  
\newcommand{\abstractfont}{\fontsize{10pt}{10pt}\selectfont} 
\newcommand{\keywordfont}{\fontsize{10pt}{10pt}\selectfont}  

\begin{document}

\title{\titlefont Data-Driven Evolutionary Game-Based Model Predictive Control for Hybrid Renewable Energy Dispatch in Autonomous Ships}

\author{%
\IEEEauthorblockN{\authorfont 1\textsuperscript{st} Yaoze Liu~${\dagger}$}
\IEEEauthorblockA{\affilfont \textit{School of Engineering, University of Strathclyde,}\\
Glasgow, United Kingdom}
\and
\IEEEauthorblockN{\authorfont 2\textsuperscript{nd} Zhen Tian~${\dagger}$}
\IEEEauthorblockA{\affilfont \textit{James Watt School of Engineering, University of Glasgow,}\\
Glasgow, United Kingdom}
\and
\IEEEauthorblockN{\authorfont 3\textsuperscript{rd} Jinming Yang}
\IEEEauthorblockA{\affilfont \textit{School of Computing Science, University of Glasgow,}\\
Glasgow, United Kingdom}
\and
\IEEEauthorblockN{\authorfont 4\textsuperscript{th} Zhihao Lin}
\IEEEauthorblockA{\affilfont \textit{James Watt School of Engineering, University of Glasgow,}\\
Glasgow, United Kingdom\\
Corresponding author: 2800400l@student.gla.ac.uk}
\thanks{\affilfont ${\dagger}$ Equal contribution}
}

\maketitle

\setlength{\parindent}{0.51cm}
\setlength{\parskip}{6pt}
\linespread{0.95}

\renewenvironment{abstract}{
  \noindent\begin{center}\bfseries Abstract\end{center}%
  \begin{quote}\abstractfont
  \setlength{\parindent}{0.48cm}\setlength{\parskip}{10pt}
}{
  \end{quote}
}

\renewenvironment{IEEEkeywords}{
  \noindent\textbf{Keywords---}\keywordfont
  \setlength{\parindent}{0.48cm}\setlength{\parskip}{6pt}
}{
}

\begin{abstract}
In this paper, we propose a data-driven Evolutionary Game-Based Model Predictive Control (EG-MPC) framework for the energy dispatch of a hybrid renewable energy system powering an autonomous ship. The system integrates solar photovoltaic and wind turbine generation with battery energy storage and diesel backup power to ensure reliable operation. Given the uncertainties in renewable generation and dynamic energy demands, an optimal dispatch strategy is crucial to minimize operational costs while maintaining system reliability.
To address these challenges, we formulate a cost minimization problem that considers both battery degradation costs and diesel fuel expenses, leveraging real-world data to enhance modeling accuracy. The EG-MPC approach integrates evolutionary game dynamics within a receding-horizon optimization framework, enabling adaptive and near-optimal control solutions in real time. Simulation results based on site-specific data demonstrate that the proposed method achieves cost-effective, reliable, and adaptive energy dispatch, outperforming conventional rule-based and standard MPC approaches, particularly under uncertainty.
\end{abstract}


\section{Introduction}
Recently, autonomous mobilizing technology has garnered significant public attention in variaous domains~\cite{lin2024enhanced,lin2024conflicts,tian2024balanced,10607945}. Among these technologies, the rapid expansion of renewable energy sources and the need for reliable and sustainable power have spurred research into autonomous ships' hybrid energy systems~\cite{diab2016novel,fei2024two,pevsa2022retrofitting}. The hybrid energy system has significant influence for remote or offshore applications, such as the autonomous ship illustrated in Fig. 1~\cite{huang2021renewable}. Offshore aquaculture farms, which rely on floating structures, require a continuous and cost-effective power supply for various operational needs, including aeration, lighting, monitoring equipment, dead fish handling system, and feeding systems~\cite{syse2016investigating,saha2022profit}. Given their offshore location, these facilities have limited access to grid electricity and often rely on diesel generators, which introduce high fuel costs and carbon emissions~\cite{jakhrani2012estimation}. Managing energy efficiently is crucial to minimize operational costs and environmental impacts~\cite{olatomiwa2016energy,vivas2018review}.
 \begin{figure}[t]
    \centering
    \includegraphics[width=0.75\linewidth]{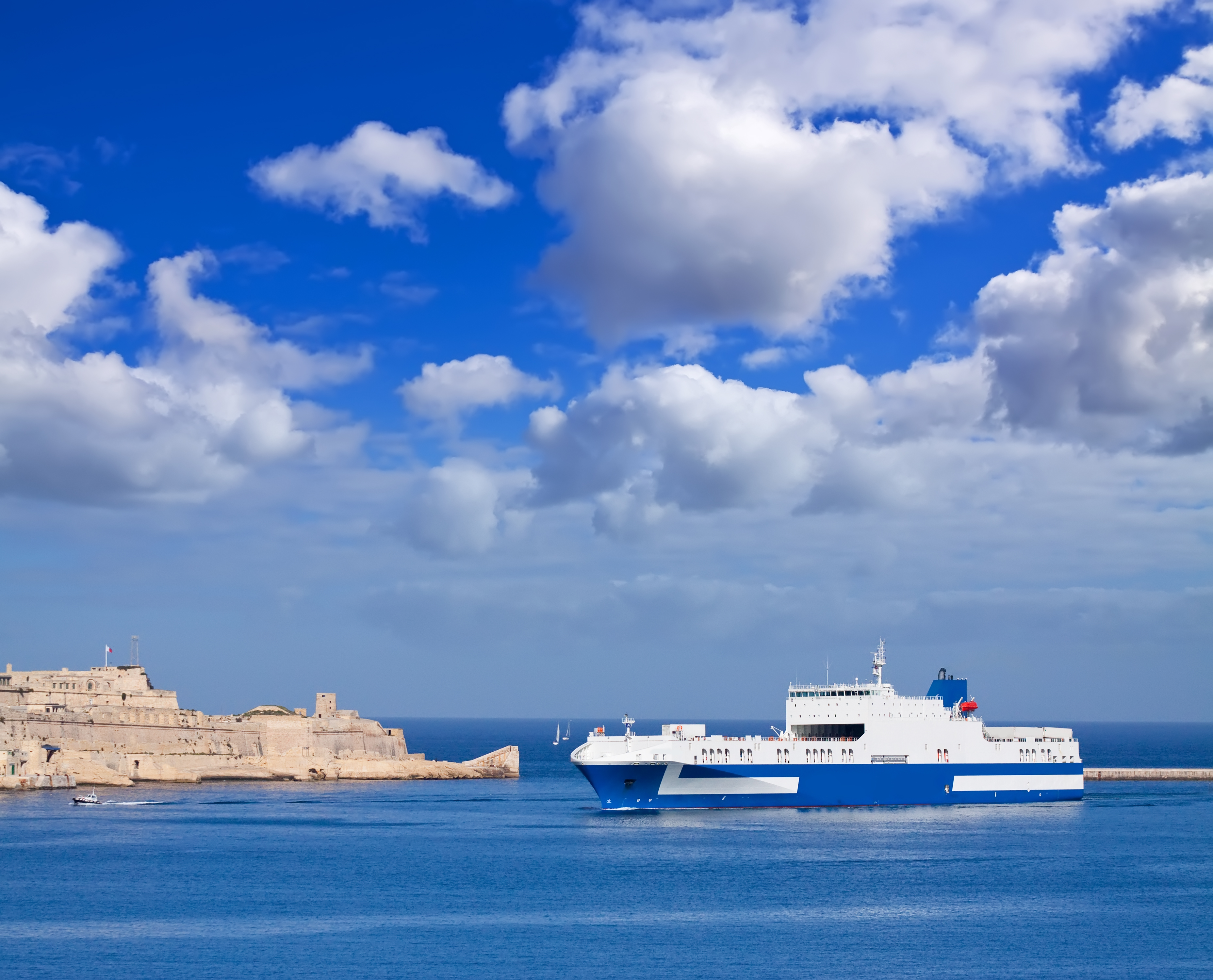}
    \vspace{-1mm}
\caption{Illustration of autonomous ship.}
    \label{fig1_framework}
    \end{figure}
     \begin{figure*}[t]
    \centering
    \includegraphics[width=0.83\linewidth]{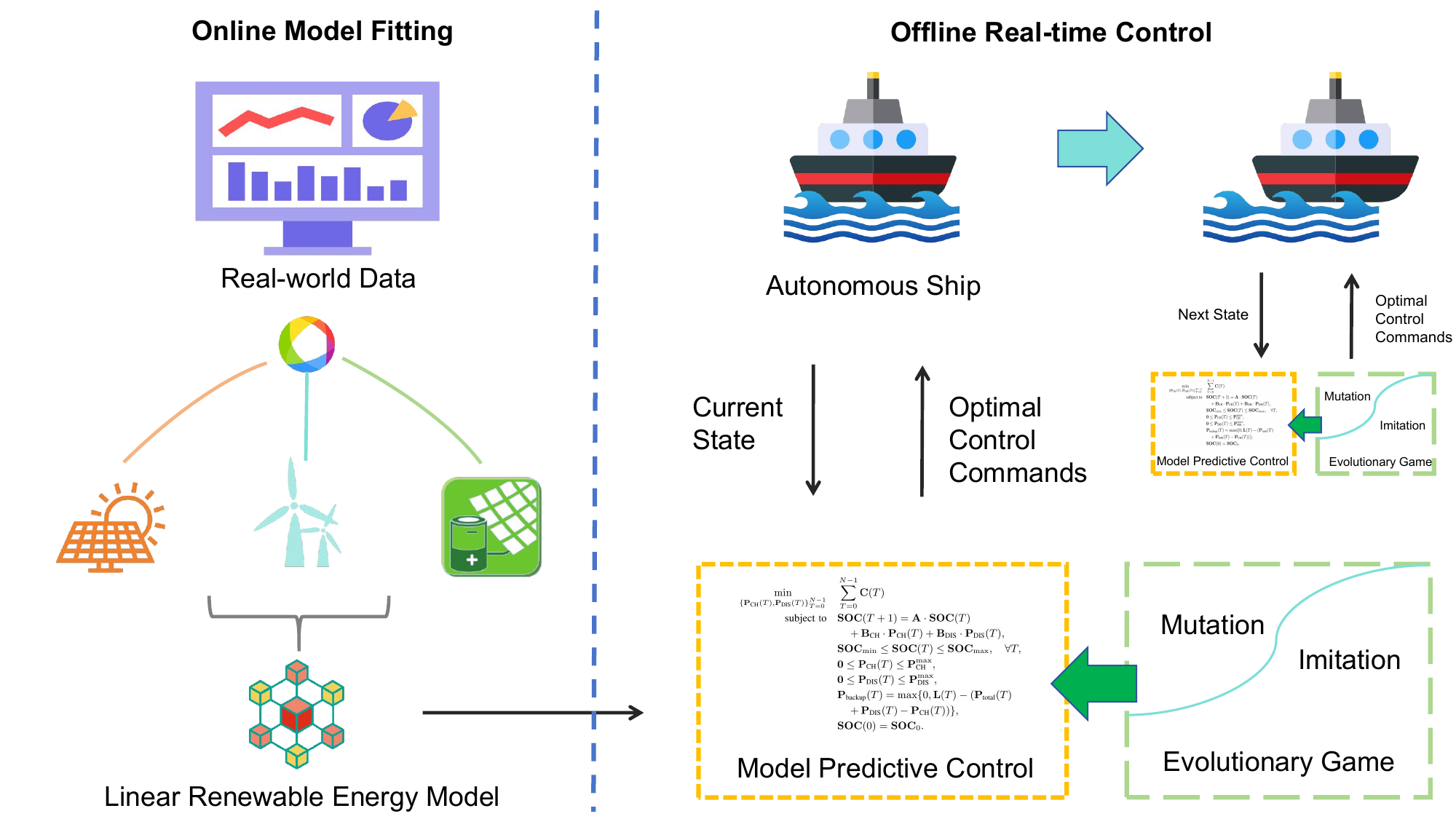}
    \vspace{-1mm}
\caption{The proposed framework using EG-MPC based on real-world data.}
    \label{fig1_framework}
    \end{figure*}
Hybrid renewable energy systems that combine solar and wind power with battery storage and diesel backup generation offer a promising solution~\cite{al2017review,roy2018electrical}. However, ensuring reliable energy supply under varying weather conditions and load demands requires sophisticated energy management strategies~\cite{olatomiwa2016energy}. Conventional rule-based optimization approaches prioritize one energy source at a time (e.g., using renewable energy first, followed by batteries, and then diesel), leading to suboptimal results~\cite{yang2022modelling}. More advanced methods such as Model Predictive Control (MPC) have been widely used for energy dispatch due to their ability to optimize multi-variable control problems with constraints~\cite{kouvaritakis2016model}. However, MPC can be computationally demanding and may struggle with uncertainty and nonlinearity in renewable generation~\cite{schwenzer2021review}.

Metaheuristic methods, such as evolutionary algorithms and swarm intelligence, have been explored to enhance the performance of MPC-based energy dispatch frameworks~\cite{mesquita2020recent,tan2013research,dokeroglu2019survey}. Traditional techniques like Ant Colony (AC) Optimization improves search efficiency but may require extensive computational resources for convergence~\cite{dorigo2007ant}. Evolutionary Game Theory (EGT), on the other hand, provides a dynamic approach using historical experience to control optimization by leveraging game-theoretic adaptation mechanisms~\cite{wang2022comprehensive}. It enables a self-adaptive search process that efficiently identifies optimal or near-optimal control actions in real-time. Given the energy cost on autonomous ships both require the robustness of MPC and the proper optimization based on historical control sequences. Therefore, the integration of evolutionary game dynamics with MPC (EG--MPC) offers a robust and computationally efficient strategy for the dispatch of a hybrid energy system dispatch under uncertainty~\cite{barreiro2017dynamical,barreiro2015evolutionary}.

This paper introduces a data-driven EG--MPC framework to optimize the energy dispatch of a hybrid renewable energy system powering an offshore aquaculture facility. Our key contributions are as follows.

\begin{enumerate}
\item We propose a linear regression-based model to characterize the hourly energy consumption of a floating aquaculture facility based on real-world data. This model captures variations in energy demand across hours.
\item We introduce an EG--MPC approach for energy dispatch, leveraging evolutionary game dynamics to enhance the search process within the receding-horizon optimization framework. This approach balances cost minimization and system reliability while adapting to fluctuating renewable energy availability.
\item We evaluate the proposed framework using site-specific data and compare its performance with conventional MPC, optimization-based strategies, and rule-based optimization strategies. The results demonstrate that EG--MPC achieves lower total costs compared to other popular benchmark algorithms.
\end{enumerate}

The remainder of this paper is structured as follows. Section 2 presents the framework of the whole system. Section 3 details the proposed EG--MPC algorithm. Section 4 discusses the simulation setup and results. Finally, Section 5 provides conclusions and future research directions.
\section{System Framework}
The proposed EG-MPC framework optimizes the hybrid renewable energy dispatch of an autonomous ship, integrating solar, wind, battery storage, and diesel backup to ensure cost-effective and reliable operation~\cite{thirunavukkarasu2023comprehensive}. The whole framework is illustrated in Fig. 2.

At its core, the MPC module provides predictive optimization over a rolling horizon, determining the best charging, discharging, and backup power usage while ensuring system constraints are met, such as battery SOC limits and renewable energy prioritization~\cite{bartolucci2019hybrid}. To enhance efficiency, an Evolutionary Game Algorithm refines dispatch strategies iteratively through selection, crossover, and mutation, reducing computational complexity and improving adaptability to uncertainties in renewable generation and energy demand.

To ensure practical applicability, the framework integrates real-world energy data from HOMER-generated synthetic resource datasets, covering solar irradiance, wind speed, and load demand at a specific maritime location. A linear regression model is used to estimate renewable energy generation based on environmental conditions, allowing aurate forecasting within the MPC structure.

The execution process follows a closed-loop optimization cycle, where the system initializes, generates candidate control sequences, evaluates cost functions, refines solutions through evolutionary game dynamics, applies the optimal control action, and updates the system state, repeating at each time step to adapt to changing conditions. This approach ensures a robust, adaptive, and computationally efficient energy management strategy that optimizes battery usage, minimizes fuel dependency, and enhances operational sustainability for autonomous ships navigating dynamic maritime environments.
\section{Methodology}

\subsection{Battery Storage Dynamics}
The battery energy storage system is modeled using a state-space representation to capture the dynamics of the state-of-charge (SOC). The SOC update equation is expressed in matrix form to handle multiple time steps efficiently. Let \(\mathbf{SOC} = [SOC(0), SOC(1), \dots, SOC(N)]^T\) represent the SOC vector over the prediction horizon \(N\). The dynamics are given by:
\begin{equation}
  \mathbf{SOC}(T+1) = \mathbf{A} \cdot \mathbf{SOC}(T) + \mathbf{B}_{\text{CH}} \cdot \mathbf{P}_{\text{CH}}(T) + \mathbf{B}_{\text{DIS}} \cdot \mathbf{P}_{\text{DIS}}(T),
  \label{eq:SOC_matrix}
\end{equation}
where \(\mathbf{A} = \mathbf{I}_{N \times N}\) is the identity matrix representing the SOC propagation, \(\mathbf{B}_{\text{CH}} = \eta_{\text{CH}} \cdot \mathbf{I}_{N \times N}\) and \(\mathbf{B}_{\text{DIS}} = -\frac{1}{\eta_{\text{DIS}}} \cdot \mathbf{I}_{N \times N}\) are the charging and discharging efficiency matrices, respectively, \(\mathbf{P}_{\text{CH}}(T) = [P_{\text{CH}}(0), P_{\text{CH}}(1), \dots, P_{\text{CH}}(N-1)]^T\) and \(\mathbf{P}_{\text{DIS}}(T) = [P_{\text{DIS}}(0), P_{\text{DIS}}(1), \dots, P_{\text{DIS}}(N-1)]^T\) are the charging and discharging power vectors.

To account for real-world data and nonlinearities, a linearized model is introduced. Let \(\Delta \mathbf{SOC}(t)\) represent the deviation from the nominal SOC trajectory. The linearized dynamics are:
\begin{equation}
\begin{split}
  \Delta \mathbf{SOC}(T+1) &= \mathbf{A} \cdot \Delta \mathbf{SOC}(T) + \mathbf{B}_{\text{CH}} \cdot \Delta \mathbf{P}_{\text{CH}}(T) \\
  &\quad + \mathbf{B}_{\text{DIS}} \cdot \Delta \mathbf{P}_{\text{DIS}}(T) + \mathbf{w}(T),
\end{split}
\label{eq:linearized_SOC}
\end{equation}
where \(\mathbf{w}(t)\) is a disturbance term capturing modeling errors and external uncertainties.

\subsection{Cost Function}
The instantaneous cost function is extended to include multiple cost components and constraints. Let \(\mathbf{C}(t) = [C(0), C(1), \dots, C(N-1)]^T\) represent the cost vector over the horizon. The cost function is formulated as:
\begin{equation}
  \mathbf{C}(t) = \mathbf{C}_{\text{bat}}(t) + \mathbf{C}_{\text{backup}}(t) + \mathbf{C}_{\text{penalty}}(t),
  \label{eq:cost_function}
\end{equation}
where \(\mathbf{C}_{\text{bat}}(t) = \mathbf{c}_{\text{bat}} \cdot \mathbf{p}_{dis}(t)\) is the battery cycling cost vector, \(\mathbf{C}_{\text{backup}}(t) = \mathbf{c}_{\text{backup}} \cdot \mathbf{p}_{\text{backup}}(t)\) is the backup generation cost vector, \(\mathbf{C}_{\text{penalty}}(t)\) is a penalty term for violating SOC constraints, defined as
 \begin{equation}
\begin{split}
  \mathbf{C}_{\text{penalty}}(T) &= \mathbf{Q} \cdot \max\{0, \mathbf{SOC}_{\min} - \mathbf{SOC}(T)\} \\
  &\quad + \mathbf{R} \cdot \max\{0, \mathbf{SOC}(T) - \mathbf{SOC}_{\max}\},
\end{split}
\label{eq:penalty}
\end{equation}
where \(\mathbf{Q}\) and \(\mathbf{R}\) are penalty coefficient matrices.

The backup power vector \(\mathbf{p}_{\text{backup}}(t)\) is computed as
\begin{equation}
  \mathbf{p}_{\text{backup}}(t) = \max\{0, \mathbf{L}(t) - (\mathbf{P}_{\text{total}}(t) + \mathbf{p}_\text{dis}(t) - \mathbf{p}_\text{ch}(t))\},
  \label{eq:backup_power}
\end{equation}
where \(\mathbf{L}(t)\) and \(\mathbf{P}_{\text{total}}(t)\) are the load and total renewable generation vectors, respectively.

\subsection{MPC Optimization Problem}
The MPC optimization problem is reformulated to handle the extended cost function and constraints. The objective is to minimize the total cost over the horizon \(N\):
{\small
\begin{equation}
\begin{split}
  \min_{\{\mathbf{P}_{\text{CH}}(T), \mathbf{P}_{\text{DIS}}(T)\}_{T=0}^{N-1}} \quad & \sum_{T=0}^{N-1} \mathbf{C}(T) \\
  \text{subject to} \quad 
  & \mathbf{SOC}(T+1) = \mathbf{A} \cdot \mathbf{SOC}(T) \\
  &\quad  + \mathbf{B}_{\text{CH}} \cdot \mathbf{P}_{\text{CH}}(T) + \mathbf{B}_{\text{DIS}} \cdot \mathbf{P}_{\text{DIS}}(T), \\
  & \mathbf{SOC}_{\min} \leq \mathbf{SOC}(T) \leq \mathbf{SOC}_{\max}, \quad \forall T, \\
  & \mathbf{0} \leq \mathbf{P}_{\text{CH}}(T) \leq \mathbf{P}_{\text{CH}}^{\max}, \\
  & \mathbf{0} \leq \mathbf{P}_{\text{DIS}}(T) \leq \mathbf{P}_{\text{DIS}}^{\max}, \\
  & \mathbf{P}_{\text{backup}}(T) = \max\{0, \mathbf{L}(T) - (\mathbf{P}_{\text{total}}(T) \\
  &\quad + \mathbf{P}_{\text{DIS}}(T) - \mathbf{P}_{\text{CH}}(T))\}, \\
  & \mathbf{SOC}(0) = \mathbf{SOC}_0.
\end{split}
\label{eq:optimization_problem}
\end{equation}
}

\subsection{Real-World Data Fitting using Linear Models}
\subsubsection{Real-World Data Fitting Process}
To effectively model the relationship between solar irradiance, wind speed, and energy generation, we fit a linear regression model to synthetic resource data generated using HOMER~\cite{ihoga_data}. The dataset spans 01/01/2018 through 31/12/2018 and corresponds to a location at Latitude: 56.4717, Longitude: -6.5521. The key parameters include Solar Irradiance [$kWh/m^2$], Wind Speed at 10m height [m/s], Farm Load Demand [kW]. Using these parameters, we establish the following linearized power generation model:
\begin{equation}
P_{renewable}(t) = \alpha_1 \cdot Irr(t) + \alpha_2 \cdot v(t) + \alpha_3 \cdot v^3(t) + \alpha_4,
\end{equation}
where $Irr(t)$ is the solar irradiance at time $t$, $v(t)$ is the wind speed, $\alpha_1, \alpha_2, \alpha_3, \alpha_4$ are coefficients estimated from regression. This model enables real-time prediction of available renewable energy, which is integrated within the MPC-based dispatch strategy to enhance energy management decisions.

\subsubsection{Load Demand and Energy Balance}
The ship's energy demand, represented as \(P_{\text{load}}(t)\), is supplied by renewable sources, battery storage, and backup generation. The energy balance equation ensures that power demand is met:
\begin{equation}
P_{\text{load}}(t) = P_{\text{renewable}}(t) + P_{\text{battery}}(t) + P_{\text{backup}}(t).
\label{eq:energy_balance}
\end{equation}
where the $P_\text{battery}$ is formulated as
\begin{equation}
    P_{\text{battery}}(t) = P_{\text{dis}}(t) - P_{\text{ch}}(t),
\end{equation}
where \( P_{\text{battery}}(t) > 0 \) means the battery is discharging, while \( P_{\text{battery}}(t) < 0 \) means the battery is charging. The $P_\text{backup}$ is formulated as
\begin{equation}
    P_{\text{backup}}(t) = \max \left( 0, P_{\text{load}}(t) - (P_{\text{renewable}}(t) + P_{\text{battery}}(t)) \right),
\end{equation}
where \( P_{\text{load}}(t) \) is the total power demand of the ship. \( P_{\text{renewable}}(t) \) is the power generated from renewable sources. \( P_{\text{backup}}(t) > 0 \) means the diesel generator is supplying power, otherwise it is off.

The system follows these operational strategies:
\begin{itemize}
\item If \(P_{\text{renewable}}(t) \geq P_{\text{load}}(t)\), excess energy is stored in the battery.
\item If \(P_{\text{renewable}}(t) < P_{\text{load}}(t)\), battery discharge or diesel backup compensates for the deficit.
\end{itemize}
\subsubsection{Rationale Selection of Linear Model}
Due to MPC solves an optimization problem at each time step. Nonlinear models increase complexity, requiring extensive computation. A linear model ensures convexity, leading to faster, real-time optimization. The ship’s energy system includes solar, wind, battery, and diesel components. While inherently nonlinear, their behavior over short time frames can be well approximated linearly, balancing accuracy and simplicity.

In addition, linear models improve control stability by keeping predictions manageable~\cite{balconi2010defence}. They also simplify enforcing operational constraints, ensuring safe battery and energy management. Solar and wind power predictions rely on empirical data. Linear regression provides accurate, interpretable, and adaptable estimations, making real-time updates feasible without excessive computation.

\subsection{MPC Optimization Formulation with Renewable Power}
To achieve optimal dispatch decisions over a prediction horizon \(N\), the MPC problem is formulated as:
\subsection{MPC Optimization Formulation with Renewable Power}
To achieve optimal dispatch decisions over a prediction horizon \(N\), the MPC problem is formulated as:
{\footnotesize
\begin{align}
  \min_{\{\mathbf{P}_{\text{CH}}(T), \mathbf{P}_{\text{DIS}}(T)\}_{T=0}^{N-1}} \quad & \sum_{T=0}^{N-1} C(T) \nonumber \\
  \text{subject to} \quad 
  & \mathbf{SOC}(T+1) = \mathbf{A} \cdot \mathbf{SOC}(T) + \mathbf{B}_{\text{CH}} \cdot \mathbf{P}_{\text{CH}}(T) \nonumber \\
  &\quad + \mathbf{B}_{\text{DIS}} \cdot \mathbf{P}_{\text{DIS}}(T) + \mathbf{W} \cdot \mathbf{P}_{\text{renewable}}(T), \nonumber \\
  & \mathbf{SOC}_{\min} \leq \mathbf{SOC}(T) \leq \mathbf{SOC}_{\max}, \quad \forall T, \nonumber \\
  & \mathbf{0} \leq \mathbf{P}_{\text{CH}}(T) \leq \mathbf{P}_{\text{CH}}^{\max}, \nonumber \\
  & \mathbf{0} \leq \mathbf{P}_{\text{DIS}}(T) \leq \mathbf{P}_{\text{DIS}}^{\max}.
\end{align}
}
\begin{algorithm}[t]
\caption{Evolutionary Game-based MPC (EG--MPC)}
\label{alg:EG_MPC}
\begin{algorithmic}[1]
    \State \textbf{Input:} Initial SOC \(\mathbf{SOC}_0\), prediction horizon \(N\), population size \(M\), mutation probability \(p_{\text{mut}}\), number of generations \(G\), vehicle and environment parameters.

    \State \textbf{Step 1: Initialize Population}
    \State Generate initial population \(\mathcal{P}\) of candidate control sequences \(\mathbf{u}_i\) for \(i = 1, \dots, M\).
    \State Define control action set \(\mathcal{A}\) using~\eqref{eq:control_actions}.

    \State \textbf{Step 2: Evaluate Candidate Sequences}
    \For{each candidate \(\mathbf{u}_i \in \mathcal{P}\)}
        \State Compute predicted SOC trajectory using~\eqref{eq:SOC_matrix}.
        \State Evaluate cost \(J(\mathbf{u}_i)\) using~\eqref{eq:cumulative_cost}.
    \EndFor

    \State \textbf{Step 3: MPC Optimization and Evolutionary Strategy}
\For{each generation \(g = 1\) to \(G\)}
    \State Select candidates for reproduction using fitness-proportionate selection.
    \State Apply crossover to generate new offspring sequences.
    \State Apply mutation with probability \(p_{\text{mut}}\) to adjust control actions.
    \State Solve MPC optimization problem using~\eqref{eq:optimization_problem}.
    \State Update population \(\mathcal{P}\) by retaining best candidates.
\EndFor

\State \textbf{Step 4: Apply Optimal Control}
\State Select the best candidate \(\mathbf{u}^*\) from \(\mathcal{P}\).
\State Apply the first control action \(u(t) = (p_{ch}^*(t), p_{dis}^*(t))\).
\State Update \(\mathbf{SOC}(t+1)\) using the system dynamics.
    \State \textbf{Return:} Optimal control sequence \(\mathbf{u}^*\)
\end{algorithmic}
\end{algorithm}

The objective function  minimizes the total cost \( \sum_{T=0}^{N-1} C(T) \), which includes battery degradation and backup generation costs. The battery dynamics are governed by the state-of-charge (SOC) update equation, which accounts for charging (\( \mathbf{P}_{\text{CH}}(T) \)), discharging (\( \mathbf{P}_{\text{DIS}}(T) \)), and the contribution of renewable power (\( \mathbf{P}_{\text{renewable}}(T) \)). The SOC is constrained within safe operating limits (\( \mathbf{SOC}_{\min} \) and \( \mathbf{SOC}_{\max} \)) to prevent overcharging or deep discharging. Additionally, the charging and discharging powers are bounded by their maximum limits (\( \mathbf{P}_{\text{CH}}^{\max} \) and \( \mathbf{P}_{\text{DIS}}^{\max} \)) to ensure safe and efficient operation. This formulation enables the system to balance energy supply and demand while leveraging renewable energy sources and maintaining battery health.
\subsection{Candidate Control Sequence Generation}
The Evolutionary Game-based MPC (EG--MPC) algorithm is enhanced to generate candidate control sequences using a more sophisticated approach. The discrete set of control actions is defined as:
\begin{equation}
  \mathcal{A} = \{ (P_{\text{CH}}, P_{\text{DIS}}) \mid P_{\text{CH}}, P_{\text{DIS}} \in \{0, \Delta P, 2\Delta P, \dots, P_{\max}\} \},
  \label{eq:control_actions}
\end{equation}
where \(\Delta p\) is the discretization step size. A candidate control sequence over the horizon \(N\) is represented as:
\begin{equation}
  \mathbf{u} = \{(P_{\text{CH}}(0), P_{\text{DIS}}(0)), \ldots, (P_{\text{CH}}(N-1), P_{\text{DIS}}(N-1))\}.
  \label{eq:candidate_sequence}
\end{equation}
\subsection{Cost Evaluation of a Candidate Sequence}
For a candidate control sequence \(\mathbf{u}\), the predicted state trajectory is computed using the battery dynamics \eqref{eq:SOC_matrix}. The cumulative cost is evaluated as:
\begin{equation}
\begin{split}
  J(\mathbf{u}) &= \sum_{T=0}^{N-1} \Bigl[ \mathbf{C}_{\text{bat}} \, P_{\text{DIS}}(T) \\
  &\quad + \mathbf{C}_{\text{backup}} \, \max\{0, L(T) \\
  &\quad  - (P_{\text{total}}(T) + P_{\text{DIS}}(T) - P_{\text{CH}}(T))\} \Bigr].
\end{split}
\label{eq:cumulative_cost}
\end{equation}

\subsection{Evolutionary Game Dynamics}
The evolutionary algorithm is extended to include advanced strategies for population evolution. The steps are as follows:
\begin{enumerate}[label=\arabic*.]
  \item \textbf{Initialization:} Generate an initial population of candidate sequences \(\{\mathbf{u}_i\}_{i=1}^{M}\) using a Latin hypercube sampling method.
  \item \textbf{Evaluation:} Compute the cost \(J(\mathbf{u}_i)\) for each candidate using \eqref{eq:cumulative_cost}.
  \item \textbf{Selection:} Use a fitness-proportionate selection mechanism to choose candidates for reproduction.
  \item \textbf{Crossover:} Apply a multi-point crossover operator to combine selected candidates and produce offspring.
  \item \textbf{Mutation:} Introduce a mutation operator that modifies elements of the candidate sequences with a probability \(p_{\text{mut}}\), selecting new actions from \(\mathcal{A}\).
  \item \textbf{Reinforcement:} Update the population by retaining the best candidates and applying a local search to refine their solutions.
\end{enumerate}

To enhance the clarity of the proposed autonomous ship dispatching technology, we supplement the paper with a flowchart that outlines the overall process of the EG--MPC framework and provide a detailed step-by-step explanation of each phase.

\begin{figure}[!t]
\centering
\begin{tikzpicture}[node distance=1.5cm, auto]
    \tikzstyle{startstop} = [rectangle, rounded corners, minimum width=3cm, minimum height=1cm, text centered, draw=black, fill=red!30]
    \tikzstyle{process}   = [rectangle, minimum width=3cm, minimum height=1cm, text centered, draw=black, fill=orange!30]
    \tikzstyle{arrow}     = [thick,->,>=stealth]
    
    \node (start) [startstop] {Start};
    \node (data) [process, below of=start] {Data Acquisition};
    \node (model) [process, below of=data] {Model Fitting};
    \node (setup) [process, below of=model] {EG--MPC Setup};
    \node (candidate) [process, below of=setup] {Candidate Generation};
    \node (evolve) [process, below of=candidate] {Evolutionary Dynamics};
    \node (optimize) [process, below of=evolve] {MPC Optimization};
    \node (control) [process, below of=optimize] {Optimal Control Action};
    \node (update) [process, below of=control] {System Update};
    \node (stop) [startstop, below of=update] {End/Next Horizon};
    
    \draw [arrow] (start) -- (data);
    \draw [arrow] (data) -- (model);
    \draw [arrow] (model) -- (setup);
    \draw [arrow] (setup) -- (candidate);
    \draw [arrow] (candidate) -- (evolve);
    \draw [arrow] (evolve) -- (optimize);
    \draw [arrow] (optimize) -- (control);
    \draw [arrow] (control) -- (update);
    \draw [arrow] (update) -- (stop);
    \draw [arrow] (update.east) -- ++(2,0) |- (data.east);
\end{tikzpicture}
\caption{Flowchart of Autonomous Ship Dispatching and the EG--MPC Process}
\label{fig:flowchart}
\end{figure}
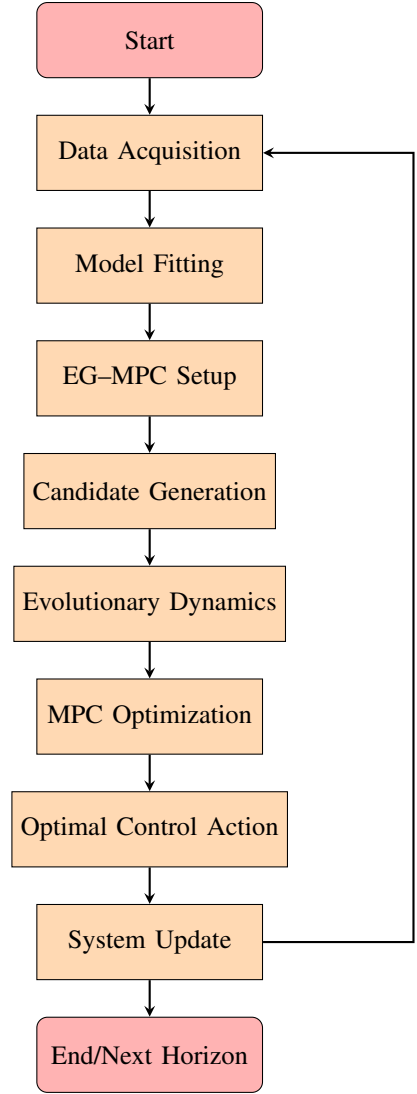
Fig.~\ref{fig:flowchart} illustrates the sequential steps of the proposed method. A detailed explanation is provided below:
\begin{figure*}[h]
    \centering
    \includegraphics[width=0.72\linewidth]{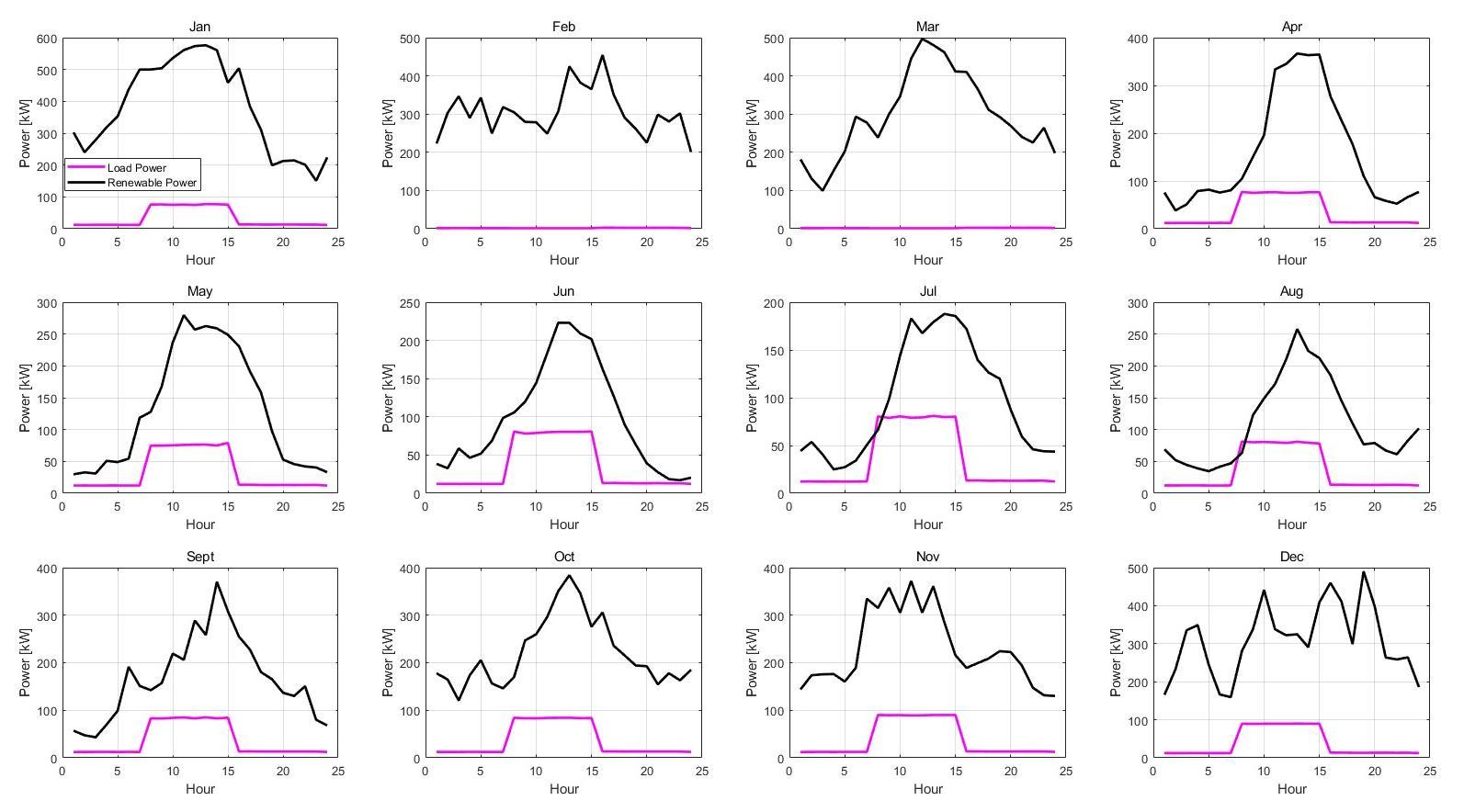}
    \vspace{-2mm}
\caption{The comparison of load power and renewable power among 12 months based on HOMER-generated synthetic resource datasets.}
    \label{fig3_lr}
    \end{figure*}

\begin{enumerate}
     \item \textbf{Data Acquisition:} The system begins by collecting all necessary data, including renewable energy profiles, such as solar irradiance and wind speed, ship load demand, and other real-world datasets. This comprehensive data collection lays the foundation for all subsequent modeling and control processes.
    
    \item \textbf{Model Fitting:} With the data in hand, a linear regression model is employed to establish the relationships between the various environmental parameters in 1). This approach provides a fast, interpretable forecasting method that is particularly well-suited for real-time applications.
    
    \item \textbf{EG--MPC Setup:} Building on the fitted model in 2), the next step is to define the MPC problem. This involves setting up the system dynamics, formulating a cost function, and establishing operational constraints. This precise definition is crucial, as it directs the optimization process toward minimizing energy costs while adhering to physical dynamics.
    
    \item \textbf{Candidate Generation:} After formulating the MPC problem by 3), an initial population of candidate control sequences is generated using a Latin hypercube sampling method. These candidates, representing discrete actions such as charging or discharging levels over the prediction horizon, provide a diverse starting point for the optimization process.
    
    \item \textbf{Evolutionary Dynamics:} The candidate control sequences in 4) are then refined using evolutionary game strategies. Techniques including fitness-proportionate selection, multi-point crossover, and mutation are applied, allowing the system to iteratively improve the candidates toward lower-cost solutions.
    
    \item \textbf{MPC Optimization:} In this phase, each refined candidate in 5) is evaluated using the cost function, and the corresponding battery state-of-charge trajectory is simulated. The candidate that results in the lowest cumulative cost is identified as the optimal control sequence.
    
    \item \textbf{Optimal Control Action:} The optimal sequence derived from 6) is then implemented on the autonomous ship’s energy system. This step ensures that the system operates with real-time responsiveness while maintaining stability.
    
    \item \textbf{System Update:} Finally, the battery state-of-charge is updated according to the battery dynamic model, and renewable energy forecasts are refreshed using the linear regression model. This update, which follows a receding-horizon strategy, prepares the system for the next control cycle, ensuring continuous and adaptive operation.
\end{enumerate}

\section{Simulation Results}
To verify the efficiency of the proposed EG-MPC, simulations are designed. This section presents our simulation results and an
analysis of the proposed framework in Matlab 2024b. In the first subsection, the linear fitting process based on HOMER-generated synthetic resource datasets is illustrated. In the second subsection, the total energy consumption using EG-MPC compared to other popular benchmarks are illustrated. The parameters used are summarized in Table. 1.
\begin{table}[h!]
\centering
\caption{Simulation Parameters and Battery Specifications}
\label{tab:parameters}
\begin{tabular}{|c|c|c|}
\hline
\textbf{Parameter} & \textbf{Value} & \textbf{Description} \\ \hline
\(T_{\text{sim}}\) & 24 & Simulation horizon (hours) \\ \hline
\(\Delta t\) & 1 & Time step (hours) \\ \hline
\(C_{\text{bat}}\) & 1000 & Battery capacity (kWh) \\ \hline
\(\text{SoC}_0\) & 500 & Initial state-of-charge (kWh) \\ \hline
\(P_{\text{ch}}^{\max}\) & 1000 & Maximum charging rate (kW) \\ \hline
\(P_{\text{dis}}^{\max}\) & 100 & Maximum discharging rate (kW) \\ \hline
\(\eta_{\text{bat}}\) & 0.9 & Charging/discharging efficiency \\ \hline
\(v(t)\) & 8 & Fixed wind speed profile (m/s) \\ \hline
\end{tabular}
\end{table}
\subsection{Linear Fitting Process}
Fig.~\ref{fig3_lr} shows the hourly load power in magenta and renewable power in black profiles for each month based on real-world data. The load power remains relatively stable throughout the day, peaking during the daytime, while renewable power varies significantly due to seasonal and environmental changes. In winter months of January, February, and December, renewable power is generally lower due to reduced solar irradiance and inconsistent wind conditions, leading to higher reliance on backup sources. In summer months of June, July, and August, renewable power peaks during daytime hours, often meeting or exceeding load demand, allowing potential energy storage for later use. Transition months of April, September, and October exhibit moderate renewable contributions with a mix of renewable and backup sources required to meet the load. The gap between renewable power and load highlights the need for efficient energy storage systems and load-shifting strategies to align demand with generation. Seasonal adjustments in energy management strategies are crucial to optimize renewable utilization and minimize reliance on backup generation~\cite{wang2015operational}.

\begin{figure}[t]
    \centering
    \includegraphics[width=1\linewidth]{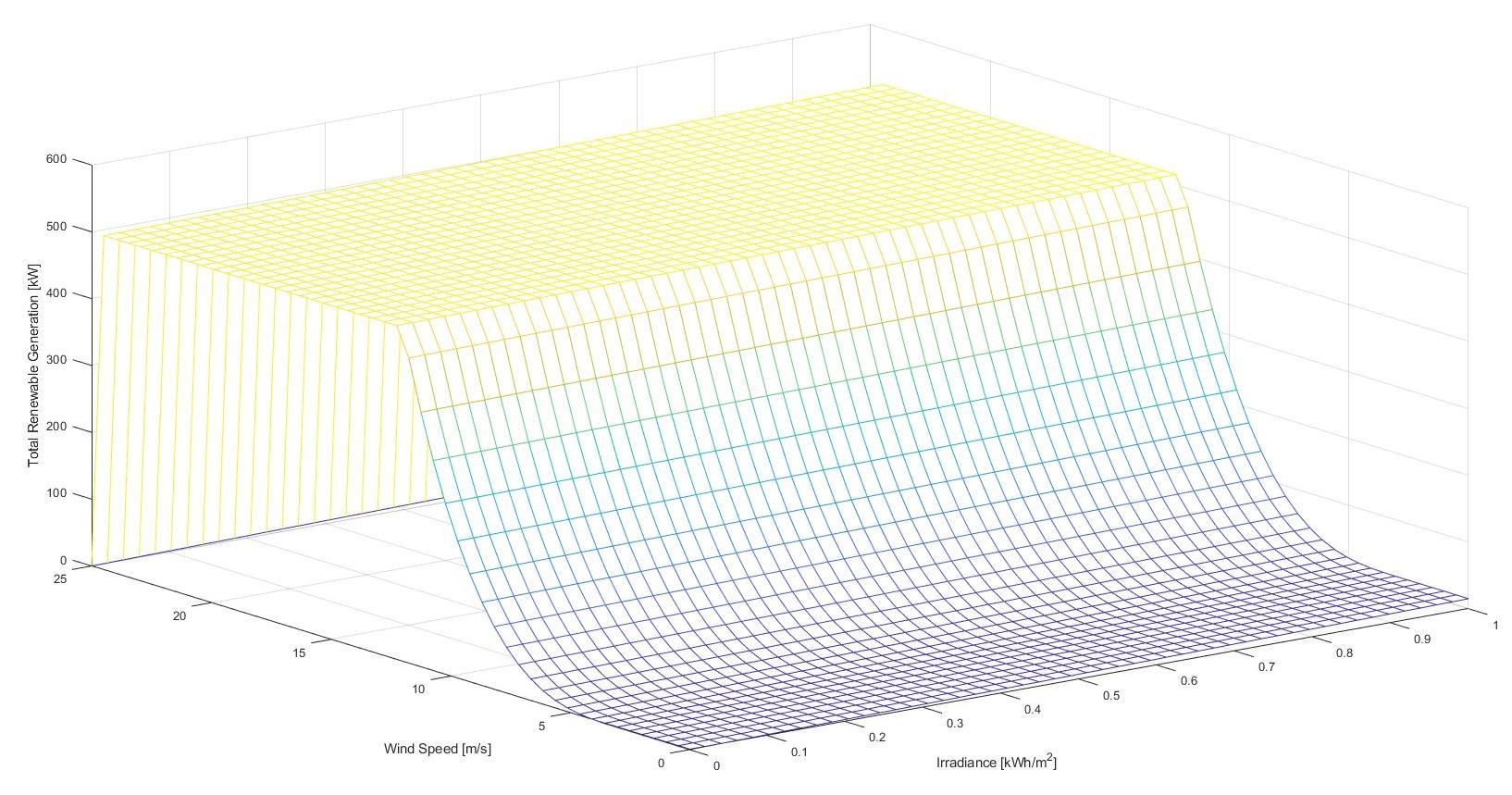}
    \vspace{-2mm}
\caption{Surface plot of total renewable energy generation as a function of wind speed and solar irradiance.}
    \label{fig4_lr}
    \end{figure}

Fig.~\ref{fig4_lr} illustrates the total renewable energy generation as a function of wind speed and solar irradiance. The plot reveals that renewable energy generation increases significantly with higher solar irradiance, reaching its peak when solar irradiance is close to 1 kWh/m\(^2\). This highlights the strong contribution of solar power under optimal conditions. Similarly, renewable energy generation shows a positive correlation with wind speed, stabilizing near maximum capacity at wind speeds above 15 m/s, where wind turbines operate at their rated power. The combined effect of solar and wind energy is evident as the total renewable output is maximized when both solar irradiance and wind speed are high. However, at low solar irradiance and low wind speeds, renewable power output drops significantly, underscoring the need for battery storage or backup power to maintain a reliable energy supply~\cite{deshmukh2008modeling}. 

    \begin{figure}[t]
    \centering
    \includegraphics[width=1\linewidth]{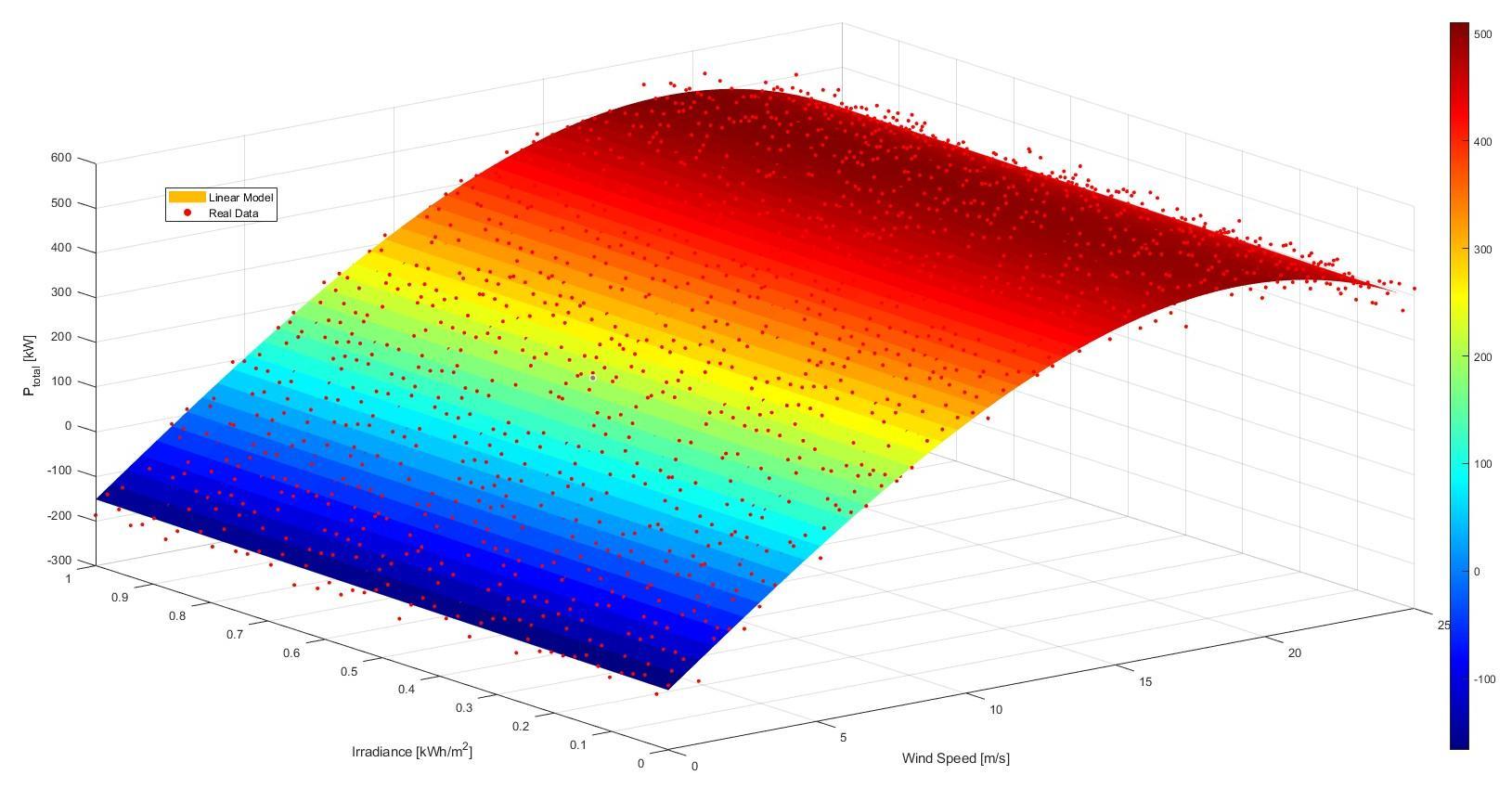}
    \vspace{-2mm}
\caption{Surface plot of linear model-generated function compared to real-data.}
    \label{fig5_lr}
    \end{figure}
The proposed linear model for renewable energy generation, \( P_{\text{total}} = -166.3272 + 15 \cdot Irr + 51.7979 \cdot v - 0.047 \cdot v^3 \), demonstrates high precision when compared to real-world data. The model effectively captures the relationship between solar irradiance, wind speed, and total renewable generation, as evidenced by its close alignment with real data points under typical environmental conditions in Fig.~\ref{fig5_lr}. 
\subsection{Effectiveness in Energy Cost of EG-MPC}
This subsection has compared the performance of EG-MPC with several popular benchmarks, including including Renewable-First, Battery-First, 50/50 Split, standard MPC, AC-MPC, and EG-MPC.
\subsubsection{Processing rules of rule-based optimization}
\textbf{1. Renewable-First Strategy:}
This strategy prioritizes renewable energy usage to meet the load demand. Any excess renewable energy is stored in the battery, and backup power is only used when both renewable and battery power are insufficient. While this minimizes renewable curtailment, it may lead to inefficient battery usage and higher reliance on backup power during low renewable generation periods~\cite{stroe2018power}.

\textbf{2. Battery-First Strategy:}
In this approach, the battery is prioritized to meet the load demand, even when renewable energy is available. Renewable energy is stored in the battery for future use, and backup power is used when the battery is depleted. This strategy often results in excessive cycling of the battery and delayed utilization of renewable energy, increasing dependency on backup power~\cite{aziz2019energy}.

\textbf{3. 50/50 Split Strategy:}
This strategy distributes energy usage equally between renewable and battery sources to meet the load demand. While it reduces over-reliance on a single source, it fails to dynamically adjust to real-time fluctuations in renewable generation and load demand, leading to suboptimal energy utilization and occasional spikes in backup power usage.
    \begin{figure}[t]
    \centering
    \includegraphics[width=0.93\linewidth]{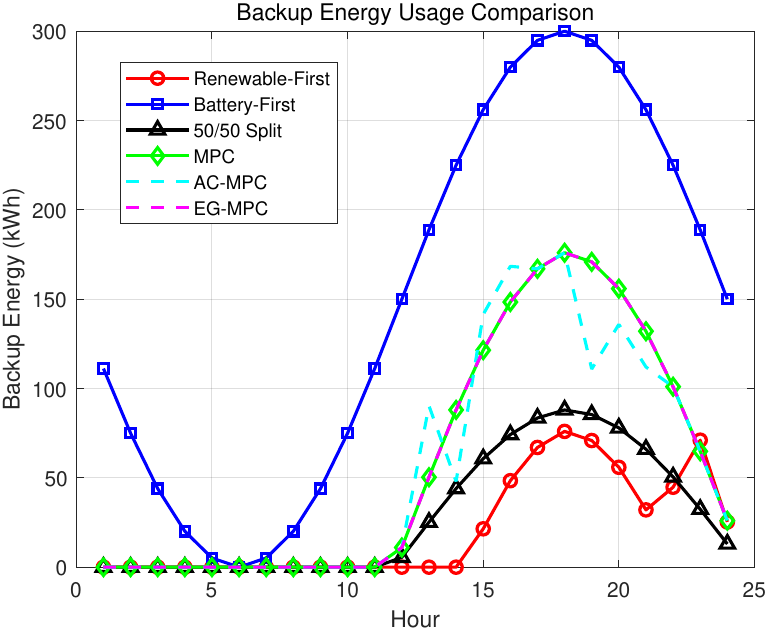}
    \vspace{-1mm}
\caption{Comparison of backup power among EG-MPC with other popular benchmark algorithms.}
    \label{fig6_lr}
    \end{figure}
\subsubsection{Comparison Analysis}
Fig.~\ref{fig6_lr} compares the backup power usage across various energy dispatch strategies, including Renewable-First, Battery-First, 50/50 Split, standard MPC, AC-MPC, and EG-MPC. The results highlight significant differences in the efficiency of each approach in minimizing backup energy consumption throughout the day. The EG-MPC strategy demonstrates superior performance by maintaining consistently low backup power usage across all hours. This is evident in its near-flat profile, indicating effective prioritization of renewable and stored energy while minimizing reliance on backup sources. In contrast, the Battery-First strategy exhibits higher and less efficient backup power usage due to its limited adaptability to real-time energy fluctuations. Specifically, the Battery-First strategy results in a sharp increase in backup usage during peak demand hours, reflecting suboptimal resource allocation.

However, after 11 hours, Renewable-First and 50/50 Split strategies show slightly lower backup power usage compared to EG-MPC. This behavior is primarily due to their static rules, which aggressively utilize renewable energy and battery storage during periods of moderate load and high renewable availability. While this reduces backup power usage temporarily, it comes at the expense of other performance metrics. The Renewable-First strategy, for example, can lead to poor energy cost performance due to inefficient battery cycling and potential renewable curtailment when the battery reaches full capacity. Similarly, the 50/50 Split strategy lacks dynamic adaptability, resulting in suboptimal energy distribution. Both strategies may cause higher battery degradation costs due to excessive cycling, as well as increased reliance on backup power during periods of fluctuating renewable generation. These factors contribute to their poor performance in long-term operational efficiency and overall energy cost minimization.
  \begin{figure}[t]
    \centering
    \includegraphics[width=1\linewidth]{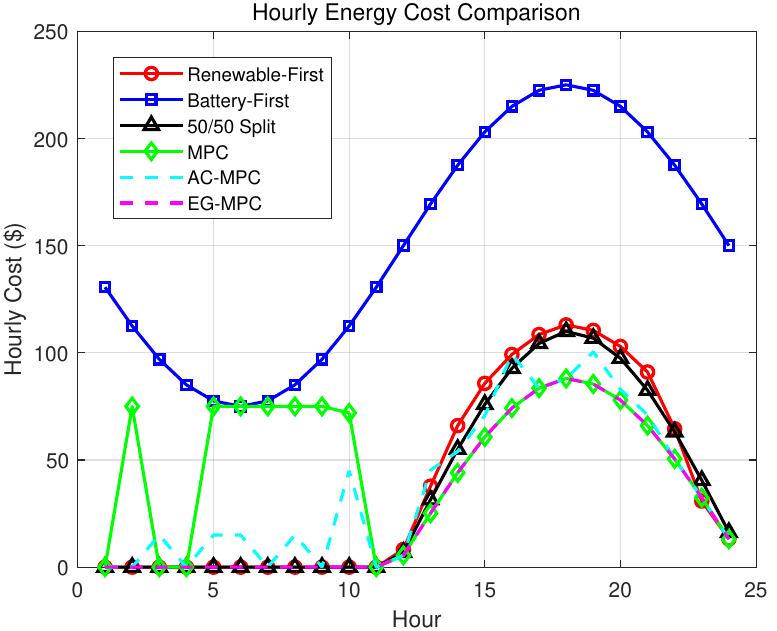}
    \vspace{-1mm}
\caption{Hourly energy cost comparison for different energy dispatch strategies.}
    \label{fig7_lr}
    \end{figure}
Fig.~\ref{fig7_lr} presents the hourly energy cost across various energy dispatch strategies, including Renewable-First, Battery-First, 50/50 Split, MPC, AC-MPC, and EG-MPC. The results highlight distinct differences in the cost-efficiency of these strategies throughout the day.

The EG-MPC approach consistently achieves the lowest energy cost over most hours, demonstrating its ability to dynamically optimize energy dispatch. By prioritizing renewable energy usage, efficiently cycling the battery, and minimizing reliance on backup power, EG-MPC ensures cost-effective operation even during high demand periods. In comparison, the Renewable-First and 50/50 Split strategies exhibit higher costs during critical peak demand hours, reflecting their limited adaptability to fluctuating load and renewable generation. While these static rule-based strategies perform adequately during hours with stable conditions, they fail to account for real-time variations, leading to suboptimal energy allocation and increased costs. The Battery-First strategy shows the highest hourly costs, especially during peak demand periods, due to excessive reliance on backup power caused by premature depletion of the battery. Similarly, standard MPC and AC-MPC strategies, although showing moderate improvements over static approaches, incur occasional cost spikes due to less effective integration of renewable energy and battery resources.
  \begin{figure}[t]
    \centering
    \includegraphics[width=1\linewidth]{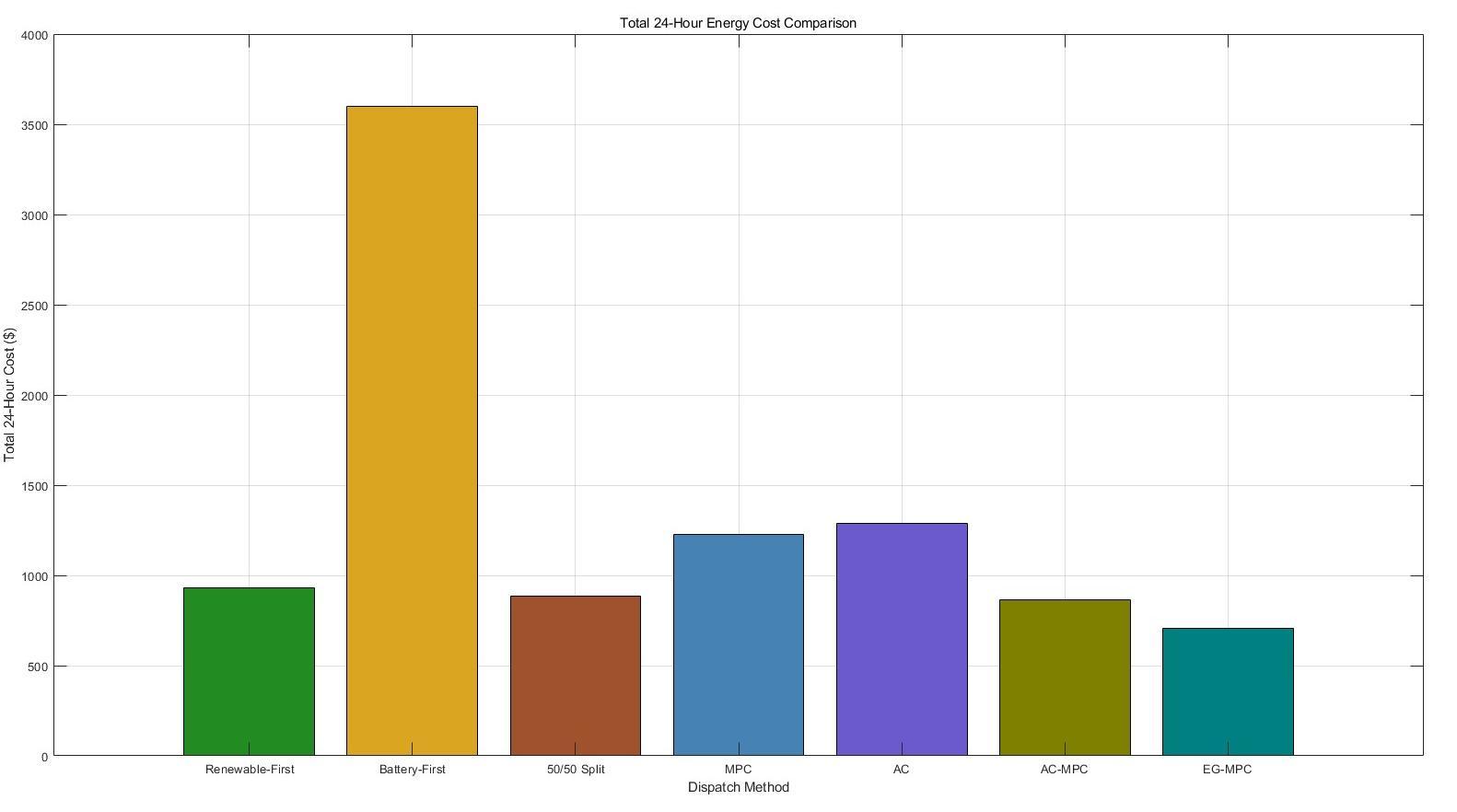}
    \vspace{-2mm}
\caption{Comparison of total cost among EG-MPC with other popular benchmark algorithms.}
    \label{fig8_lr}
    \end{figure}
    
Fig.~\ref{fig8_lr} compares the total 24-hour energy costs across various energy dispatch strategies, including Renewable-First, Battery-First, 50/50 Split, MPC, AC-MPC, and EG-MPC. The results highlight significant differences in the cost-effectiveness of these strategies.

The EG-MPC achieves the lowest total energy cost, demonstrating its superior ability to dynamically optimize energy dispatch by efficiently balancing renewable energy utilization, battery cycling, and backup power. Its adaptability to real-time conditions minimizes unnecessary energy expenditures, making it the most cost-efficient approach. In contrast, the Battery-First strategy incurs the highest total cost due to its inefficient energy allocation. By prioritizing the battery for load demand, this strategy prematurely depletes battery storage, leading to excessive reliance on costly backup power. The 50/50 Split strategy also shows suboptimal performance, as its fixed allocation fails to adjust to fluctuating renewable generation and load demand, resulting in elevated costs.

The Renewable-First strategy performs better than Battery-First and 50/50 Split, benefiting from prioritizing renewable energy usage. However, it still incurs higher costs than EG-MPC due to occasional inefficiencies in battery usage and backup reliance during low renewable generation periods. The standard MPC and AC-MPC strategies improve upon static rule-based methods but fall short of EG-MPC due to their limited ability to handle dynamic and complex energy dispatch scenarios.

\section{Conclusion}
This work addresses the critical challenge of optimizing energy dispatch in hybrid renewable energy systems for autonomous applications, focusing on minimizing costs while ensuring reliability. The importance of this problem lies in the increasing reliance on renewable energy sources and the need to reduce dependency on non-renewable backup power, which is both costly and environmentally unsustainable. To tackle this issue, we proposed a novel EG-MPC framework. The rationale behind EG-MPC lies in its ability to dynamically optimize energy dispatch in real-time, leveraging evolutionary game dynamics to efficiently search the solution space and adapt to changing conditions~\cite{barreiro2015evolutionary,tanimoto2015fundamentals}. A key component of our approach is the integration of a linear model derived from real-world data, capturing the relationship between renewable energy generation, solar irradiance, and wind speed. This linear model enables computational efficiency and accurate predictions, forming the foundation of our optimization framework. The simulation results demonstrated the effectiveness of EG-MPC across multiple scenarios. Using real-world data, the linear model was shown to closely approximate renewable energy generation, providing reliable input for the control strategy. Compared to benchmark strategies, including Renewable-First, Battery-First, 50/50 Split, MPC, and AC-MPC, EG-MPC achieved the lowest energy costs while minimizing backup power usage. The 24-hour total cost analysis further highlighted its superiority, outperforming static and less adaptive approaches by dynamically balancing renewable utilization, battery cycling, and backup reliance. Further works could be building up the cooperated energy strategies for connected autonomous ships based on current EG-MPC for single autonomous ship.

\bibliographystyle{IEEEtran}
\bibliography{IEEEabrv,zq_lib}

\end{document}